\documentclass[11pt]{article}
\usepackage{amsmath,amssymb,color,epsf}

\makeatletter
\@addtoreset{equation}{section}
\makeatother

\newcommand{\hoch}[1]{$\, ^{#1}$}

\newcommand{\be}{\begin{equation}}
\newcommand{\ee}{\end{equation}}
\newcommand{\bea}{\setlength\arraycolsep{2pt} \begin{eqnarray}}
\newcommand{\eea}{\end{eqnarray}}
\newcommand{\nn}{\nonumber}

\def\ft#1#2{{\textstyle{\frac{\scriptstyle #1}{\scriptstyle #2} } }}
\def\fft#1#2{{\frac{#1}{#2}}}

\def\0{{\sst{(0)}}}
\def\1{{\sst{(1)}}}
\def\2{{\sst{(2)}}}
\def\3{{\sst{(3)}}}
\def\4{{\sst{(4)}}}
\def\5{{\sst{(5)}}}
\def\6{{\sst{(6)}}}
\def\7{{\sst{(7)}}}
\def\8{{\sst{(8)}}}
\def\sst#1{{\scriptscriptstyle #1}}

\def\tA{{\widetilde A}}
\def\tP{{\widetilde P}}

\begin{document}

\begin{flushright}
\hfill{ \
MIFPA-12-34\ \ \ \ }
\end{flushright}

\begin{center}
{\Large {\bf Supersymmetric Solutions in Four-Dimensional Off-Shell
Curvature-Squared Supergravity}
}

\vspace{10pt}

{\Large
Hai-Shan Liu\hoch{1}, H. L\"u\hoch{2}, Yi Pang\hoch{3}, C.N. Pope\hoch{3,4}
}

\vspace{10pt}

\hoch{1} {\it Institute for Advanced Physics \& Mathematics,
Zhejiang University of Technology, Hangzhou 310032, China}

\vspace{10pt}

\hoch{2}{\it Department of Physics, Beijing Normal University,
Beijing 100875, China}

\vspace{10pt}

\hoch{3} {\it George P. \& Cynthia Woods Mitchell  Institute
for Fundamental Physics and Astronomy,\\
Texas A\&M University, College Station, TX 77843, USA}

\vspace{10pt}

\hoch{4}{\it DAMTP, Centre for Mathematical Sciences,
 Cambridge University,\\  Wilberforce Road, Cambridge CB3 OWA, UK}

\vspace{20pt}

\underline{ABSTRACT}
\end{center}

   Off-shell formulations of supergravities allow one to add
closed-form
higher-derivative super-invariants that are separately supersymmetric
to the usual lower-derivative actions. In this paper we study
four-dimensional off-shell ${\cal N}=1$ supergravity where additional
super-invariants associated with the square of the Weyl tensor and the
square of the Ricci scalar are included. We obtain a variety of solutions
where the metric describes domain walls, Lifshitz geometries, and also solutions
of a kind known as gyratons. We find that in some cases the solutions can
be supersymmetric for appropriate choices of the parameters. In some
solutions the auxiliary fields may be imaginary.  One may reinterpret
these as real solutions in an analytically-continued theory.  Since the
supersymmetry transformation rules now require the gravitino to be complex,
the analytically-continued theory has a ``fake supersymmetry'' rather than
a genuine supersymmetry.  Nevertheless, the concept of pseudo-supersymmetric
solutions is a useful one, since the Killing spinor equations provide
first-order equations for the bosonic fields.

\vspace{15pt}

\thispagestyle{empty}

\pagebreak
\voffset=-40pt
\setcounter{page}{1}

\tableofcontents

\addtocontents{toc}{\protect\setcounter{tocdepth}{2}}


\newpage

\section{Introduction}

   The study of supersymmetric solutions in theories of supergravity has
proved to be extremely fruitful over the years.  Much of the focus has been
on those supergravities that are directly related to string theory or
M-theory, and mostly at the level of the leading-order theories, such
as $D=11$ supergravity or the type IIA and IIB supergravities in
$D=10$.  It is known that in string theory or M-theory these supergravities
will receive higher-order corrections, including, in particular, terms in
the effective actions involving higher powers of the curvature tensor. In
fact, these corrections are expected to continue to arbitrarily high order
in powers of the curvature.  In general, it is inevitable that once
any higher-order terms are included in the ten or eleven-dimensional
action, the process of supersymmetrising them will be an endless one, requiring
corrections to the action and to the transformation rules at all orders.

In dimensions $D\le 6$, things can be rather different, because in these cases
there exist off-shell formulations of certain supergravity theories.  In such
cases, the possibility arises of being able to add a finite number of
higher-order terms to an existing supersymmetric action, in the form of
complete and self-contained super-invariants, such that the resulting theory
is fully supersymmetric in its own right, and with no modifications to
the original transformation rules.  Such theories can provide
interesting insights into the effects of higher-order curvature terms on the
solutions of the theories, while retaining the advantages of a theory
that is self-contained and allowing the possibility of obtaining corrected
solutions in closed form.

In earlier papers \cite{luwangsusylif,lupogyraton}, solutions were investigated in the four-dimensional off-shell ${\cal N}=1$ supersymmetrisation of
Einstein-Weyl gravity.  Four-dimensional ${\cal N}=1$ off-shell
supergravities \cite{sw,fv} were recently used to obtain theories with rigid
supersymmetry, by taking a limit in which gravity decoupled \cite{fessei}.
The theory considered in \cite{luwangsusylif,lupogyraton} involved
one higher-order curvature invariant, namely the Weyl-squared term.
Einstein-Weyl gravity was shown to have a critical point, for a specific
choice of the coefficient of the Weyl-squared term,
where the massive graviton disappears and is replaced by a spin-2
mode with a logarithmic fall off \cite{lpcritical}.
It is a four-dimensional generalization of chiral gravity in
three dimensions \cite{chiralgrav}.  It turns out that the theory
can be supersymmetrised in the off-shell formalism.  The critical
behaviour of the resulting Einstein-Weyl supergravity was
studied in \cite{lpsw}.

  The focus in \cite{luwangsusylif,lupogyraton} was on solutions having the
form of Lifshitz or gyrating Schr\"odinger geometries.
Amongst the solutions obtained there were supersymmetric examples.
There are in fact two independent
off-shell superinvariants involving quadratic curvature terms that can be
added in the ${\cal N}=1$ theory.  One of these involves only the
square of the Weyl tensor; this is the one that was considered in
\cite{luwangsusylif,lupogyraton}.  The other quadratic invariant involves
only the square of the Ricci scalar.  In the present paper,
we shall look for
supersymmetric solutions, but within the wider class of ${\cal N}=1$ off-shell
supergravities involving both of these quadratic curvature invariants.

    The paper is organized as follows. In section 2, we review the
four-dimensional ${\cal N}=1$ off-shell supergravity, including all the
super-invariants involving powers of curvature ranging from zero to four.
For our purposes, it suffices to present only the bosonic Lagrangian and
the supersymmetry transformation rule for the gravitino.
Owing to the global symmetry of certain of the super-invariants and the
way the scalar auxiliary fields $S$ and $P$ couple with the
auxiliary vector field $A_\mu$,
it is convenient to refer to the off-shell supergravity as the $U(1)$ theory.
Motivated by some of the solutions we obtain, it is natural also to consider
an analytically-continued theory in which the original
auxiliary pseudoscalar and vector fields $P$ and $A_\mu$ become imaginary.
The resulting bosonic Lagrangian remains real, but the supersymmetry
transformation rules require the gravitino to be complex, implying that
the theory, which we refer to as the $O(1,1)$ theory, is actually a
``fake supergravity.''  In section 3, we find domain wall solutions
supported by the auxiliary scalars and/or the vector, and we study
their supersymmetry.  In the process, we obtain all the supersymmetric
AdS vacua. Both supersymmetric singular domain walls and wormholes can arise
in these higher-order off-shell supergravities.

     In section 4 we consider Lifshitz solutions and list all possible
homogeneous Lifshitz vacua utilising scalar and/or vector auxiliary
fields.  We find that Killing spinors can arise for suitable choices
of parameters in the Lifshitz
solutions, but only in the case of the analytically-continued
$O(1,1)$ theory.  These solutions are therefore pseudo-supersymmetric in the
$O(1,1)$ fake supergravity.  In section 5, we obtain pseudo-supersymmetric
asymptotically Lifshitz solutions in the $O(1,1)$ theory.  In section 6,
we consider homogeneous gyrating Sch\"odinger vacua in both the
$U(1)$ and the $O(1,1)$ theories, and we tabulate the general solutions.
We then look for parameter choices giving rise to Killing spinors,
and find that these arise only in the $U(1)$ theory, implying the
existence of supersymmetric gyrating solutions.
We extend the discussion in section 7, to consider a more general class
of gyrating pp-wave solutions, amongst which we find a large class of
supersymmetric solutions.  The paper ends with conclusions in section 8.

\section{${\cal N}=1$, $D=4$ off-shell supergravity}

The field content of off-shell ${\cal N}=1$, $D=4$ supergravity
comprises the metric $e_{\mu}^a$, a massive vector $A_\mu$ and a
complex scalar ${\cal M}=S + {\rm i} P$, totalling 12 off-shell
degrees of freedom, matching with that of the off-shell gravitino
$\psi_\mu$. The general formalism for constructing a supersymmetric
action for any chiral superfield was presented in
\cite{Binetruy:2000zx}. For appropriate choices of superfields, one
obtains the actions of the supersymmetrisations of the cosmological
term, the Einstein-Hilbert term and higher-order curvature terms.

   The supersymmetrisation of the Einstein-Hilbert term was obtained
in \cite{sw,fv}.  In this theory, the complex scalar and the
massive vector are both auxiliary, with purely algebraic equations
of motion.  These fields can be integrated out, giving rise to standard
on-shell ${\cal N}=1$, $D=4$ supergravity.  One defining property of off-shell
supergravity is that the supersymmetry transformation rules close
without needing to make use of the equations of motion. This
implies that one may construct new theories by adding additional
super-invariants, which can in general involve higher-derivatives,
without any modification to the supersymmetry transformation rules. It turns
out that there are two quadratic curvature super-invariants: One is
the Weyl-squared super-invariant and the other is the $R^2$ super-invariant
\cite{ledu}.  In the former case, the scalars $S$ and $P$ remain
auxiliary whilst the vector $A_\mu$ acquires a kinetic term. In the latter
case, the scalars acquire derivative terms as well.  (Note that we 
shall continue to refer to $S+ {\rm i}P$ and $A_\mu$ as auxiliary fields, 
even though they start to propagate after the higher-order super-invariants 
are included.) For our purposes, we shall present only the bosonic 
Lagrangian and the supersymmetry
transformation rule for the gravitino.

\subsection{The $U(1)$ supergravity theory}

Together with the supersymmetrisation of the cosmological term
\cite{lpsw}, the bosonic action is given by
\begin{equation}
I=\int d^4 x \sqrt{-g}\, \Big(\sigma {\cal L}_0 + \lambda {\cal
L}_{S} + \ft12\alpha\, {\cal L}_C + \ft12\beta\, {\cal
L}_{R^2}\Big)\,,
\end{equation}
where $\sigma$, $\lambda$, $\alpha$ and $\beta$ are constants, and
\begin{eqnarray}
{\cal L}_0 &=& R - \ft23 ({\cal M}\bar {\cal M} - A^2)\,, \qquad
{\cal M}=S+ {\rm i} P\,,\cr {\cal L}_S &=& {\cal M} + \bar{\cal
M}\,,\cr {\cal L}_C &=& C_{\mu\nu\rho\sigma} C^{\mu\nu\rho\sigma} -
\ft23 F^2\,,\cr {\cal L}_{R^2} &=& R^2 + \ft43 (A^2 + \ft12 {\cal
M}\bar {\cal M}) R + 4 (\nabla_\mu A^\mu)^2 -4 \partial_\mu {\cal M}
\partial^\mu \bar {\cal M}\cr && -\ft43 {\rm i} A^\mu (\bar {\cal M}
\partial_{\mu} {\cal M} - {\cal M} \partial_\mu \bar {\cal M}) +
\ft{4}{9} ( {\cal M}^2\bar {\cal M}^2 + {\cal M}\bar {\cal M} A^2 +
A^4)\cr &=& R^2 + \ft43 (A^2 + \ft12 {\cal M}\bar {\cal M}) R + 4
(\nabla_\mu A^\mu)^2 - 4 |D_\mu {\cal M}|^2 + \ft49 (|{\cal M}|^2 +
A_\mu A^\mu)^2\,,\label{u(1)theory}
\end{eqnarray}
where
\begin{eqnarray}
F_{\mu\nu} = \partial_\mu A_\nu - \partial_\nu A_\mu\,,\qquad
A^2=A^\mu A_\mu\,,\qquad F^2 = F^{\mu\nu} F_{\mu\nu}\,,\cr
D_\mu{\cal M}=\partial_\mu {\cal M} - \ft13{\rm i} A_\mu {\cal
M}\,,\qquad D_\mu\bar {\cal M}=\partial_\mu \bar {\cal M} +\ft13{\rm
i} A_\mu \bar {\cal M}\,.\label{variousdef1}
\end{eqnarray}
The ${\cal L}_C$ and ${\cal L}_{R^2}$ super-invariants were given 
in \cite{ledu}.
The constant $\sigma$ in general can be set to 1 by appropriate scalings.
However, it is convenient here to allow it to remain
arbitrary, to emphasize that the terms
${\cal L}_0$, ${\cal L}_S$, ${\cal L}_C$ and ${\cal L}_{R^2}$ are
independent super-invariants, and each can be turned on or off
independently. The supersymmetry transformation rule for the
gravitino is universal, and given by
\begin{equation}
\delta \psi_\mu = - D_\mu \epsilon - \ft{\rm i}{6} (2A_\mu -
\Gamma_{\mu\nu} A^\nu) \Gamma_5 \epsilon - \ft16 \Gamma_\mu (S +
{\rm i} \Gamma_5 P)\epsilon\,.
\end{equation}

The equation of motion for the complex scalar ${\cal M}$ is given by
\begin{equation}
-\ft23\sigma {\cal M} + \lambda + \ft12\beta \Big(\ft23 {\cal M} R +
4\Box {\cal M}- \ft43 {\rm i} (2A^\mu \partial _\mu {\cal M} + {\cal
M}\nabla_\mu A^\mu ) + \ft49 {\cal M} (2 {\cal M}\bar {\cal M} +
A_\mu A^\mu)\Big)=0\,,\label{speom1}
\end{equation}
and the equation of motion for the vector $A_\mu$ is given by
\begin{eqnarray}
0&=&\ft23\alpha \nabla_\mu F^{\mu\nu} + \ft23\sigma A^\nu +
\beta\Big( \ft23R\,A^\nu - 2\nabla^\nu (\nabla_\mu A^\mu) - \ft13
{\rm i} (\bar {\cal M}\nabla^\nu {\cal M} - {\cal M}\nabla^\nu \bar
{\cal M})\cr &&\qquad\qquad\qquad\qquad\qquad +\ft29 {\cal M}\bar
{\cal M} A^\nu + \ft49 A^2 A^\nu\Big)\,.
\end{eqnarray}
The Einstein equation of motion is
\begin{equation}
\sigma E^0_{\mu\nu} + \lambda E^{S}_{\mu\nu} + \alpha E^C_{\mu\nu} +
\beta E^{R^2}_{\mu\nu}=0\,,
\end{equation}
where
\begin{eqnarray}
E^0_{\mu\nu} &=& R_{\mu\nu}- \ft12 R g_{\mu\nu} + \ft13 g_{\mu\nu}
{\cal M} \bar {\cal M}\, g_{\mu\nu} +\ft23 (A_\mu A_\nu - \ft12 A^2
g_{\mu\nu}) \,,\cr 
E^{S}_{\mu\nu}&=&-\ft12 g_{\mu\nu} ({\cal M} + \bar {\cal M})\,, \cr
E_{\mu\nu}^C &=& -(2\nabla^\rho\nabla^\sigma +
R^{\rho\sigma})C_{\mu\rho\sigma\nu} - \ft23
(F_{\mu\nu}^2 - \ft14 F^2 g_{\mu\nu})\,,\cr %
E^{R^2}_{\mu\nu} &=& 2 R R_{\mu\nu} - 2\nabla_\mu\nabla_\nu R +
  2\square R\, g_{\mu\nu} + \ft43 A_\mu A_\nu R+\cr
&& \ft43(R_{\mu\nu} -\nabla_\mu\nabla_\nu + g_{\mu\nu} \square) (A^2
+ \ft12 {\cal M} \bar{\cal M})+ 4 g_{\mu\nu} \nabla_\rho(A^\rho\,
\nabla_\sigma A^\sigma)\cr 
&& -
 8 A_{(\mu} \nabla_{\nu)} \nabla_\rho A^\rho -
  4 D_{(\mu} {\cal M} \, D_{\nu)}\bar{\cal M} +
  \ft89 A_\mu A_\nu ({\cal M}\bar{\cal M} + A^2)\cr
&& - \ft12 g_{\mu\nu}\Big[ R^2 + \ft43 (A^2+ \ft12 {\cal M}\bar{\cal
M}) R + 4 (\nabla_\rho A^\rho)^2\cr &&\qquad\quad - 4 D_\rho {\cal
M}\, D^\rho\bar{\cal M} + \ft49 ({\cal M}\bar{\cal M} +
A^2)^2\Big]\,.
\end{eqnarray}
Note that the derivatives of the complex scalar $M$ that appear in
the action, $D_\mu {\cal M}$ and $D_\mu \bar {\cal M}$, are defined
in (\ref{variousdef1}).  Thus the complex scalar can be viewed as being
``charged'' under the $U(1)$ vector.  Furthermore,  if we set
$\lambda=0$ the complex scalar has a $U(1)$ global symmetry ${\cal
M}\rightarrow e^{{\rm i} \theta} {\cal M}$. For this reason, we shall refer to
this action as the $U(1)$ theory.

\subsection{The $O(1,1)$ ``fake supergravity'' theory}

    If we perform the field
redefinitions
\be
A_\mu = {\rm i} \tA_\mu\,,\qquad P =
-{\rm i} \tP\,,\label{redef}
\ee
where $\tA_\mu$ and $\tP$ are taken to be real,
the bosonic Lagrangian remains real.  We now have
\begin{eqnarray}
{\cal L}_0 &=& R - \ft23 ({\cal M}\check {\cal M} + \tA^2)\,, \qquad
{\cal M}=S+ \tP\,,\qquad \check M=S-\tP\cr {\cal L}_S &=& {\cal M} +
\check{\cal M}\,,\cr {\cal L}_C &=& C_{\mu\nu\rho\sigma}
C^{\mu\nu\rho\sigma} + \ft23 F^2\,,\cr
{\cal L}_{R^2} &=& R^2 +
\ft43 (\ft12 {\cal M}\check {\cal M}-\tA^2) R - 4 (\nabla_\mu \tA^\mu)^2
-4
\partial_\mu {\cal M}
\partial^\mu \check {\cal M}\cr && +\ft43 \tA^\mu (\check {\cal M}
\partial_{\mu} {\cal M} - {\cal M} \partial_\mu \check {\cal M}) +
\ft{4}{9} ( {\cal M}^2\check {\cal M}^2 - {\cal M}\check {\cal M} \tA^2 +
\tA^4)\label{o11theory}\\
&=& R^2 + \ft43 (\ft12 {\cal M}\check {\cal M}-\tA^2) R - 4
(\nabla_\mu \tA^\mu)^2
- 4 D_\mu {\cal M} D_\mu \check {\cal M} +
 \ft49 ({\cal M}\check{\cal M} - \tA^2 )^2\,,\nn
\end{eqnarray}
where
\begin{equation}
D_\mu{\cal M}=\partial_\mu {\cal M} + \ft13 \tA_\mu {\cal M}\,,\qquad
D_\mu\check {\cal M}=\partial_\mu \check {\cal M} - \ft13\tA_\mu \check
{\cal M}\,,
\end{equation}
and $\sigma$, $\lambda$, $\alpha$ and $\beta$ are constants.
Thus we see that the scalars are gauged in the original Weyl
sense.  The Lagrangian, if we set $\lambda=0$, is invariant under
a $O(1,1)$ global symmetry that acts as a boost on the scalars $S$ and
$\tP$.  We shall refer to this theory as the $O(1,1)$ theory.

   The analytic continuation of the $A_\mu$ and $P$ fields can be thought
of as a choice of a different ``real section'' of the complexification of
the original theory.  The process of complexifying a supergravity theory was
discussed in detail in \cite{behaplrovd}.  If one first writes the theory
in terms of purely holomorphic functions of the original real variables (in
particular, in the fermionic sector, in terms of Majorana spinors with
all conjugations being performed using the Majorana rather than the Dirac
conjugate), then almost trivially, the theory remains supersymmetric if
all the real fields are now allowed to become complex.  Of course, the
action will now be complex also, and the numbers of bosonic and fermion
degrees of freedom will be doubled.  The question then arises as to whether
there exist alternative possibilities for choosing real sections, by
imposing appropriate conjugation conditions on all the fields, such that one
again obtains a real action and a consistent set of supersymmetry
transformation rules for a genuine supergravity theory.  Finding a consistent
choice of conjugation conditions on the bosonic fields that results again in
a real bosonic action is a necessary part of this procedure.  If one can at
the same time also impose a set of conjugation conditions on the fermionic
fields such that their action is real and the supersymmetry transformations
are consistent with the conjugation properties, then one has arrived at a
genuine supergravity.  If, on the other hand, it is not possible to
impose such conjugation conditions on the fermions, then one has instead
arrived at a ``fake supergravity,'' meaning in particular that the fermions
are necessarily complex rather than being purely real or purely imaginary.

   In the present case of interest, it turns out that having imposed
our conjugation conditions $A_\mu^* = -A_\mu$ and $P^*=-P$ on the original,
but now complexified, $A_\mu$ and $P$ fields, it is not possible to find
a consistent choice of conjugation section of the complexified fermion
fields that
halves their degrees of freedom again.  They must necessarily remain
complex, and so the $O(1,1)$ theory is therefore a ``fake supergravity." It
is still of interest, however, since it provides us with
a real bosonic theory that admits real  bosonic ``pseudo-supersymmetric''
solutions that obey first-order equations following from the
requirement of the existence of complex pseudo-Killing spinors.

The
pseudo-supersymmetry transformation rule for the off-shell gravitino is now
given by
\begin{equation}
\delta \psi_\mu = - D_\mu \epsilon + \ft{1}{6} (2\tA_\mu -
\Gamma_{\mu\nu} \tA^\nu) \Gamma_5 \epsilon - \ft16 \Gamma_\mu (S +
\Gamma_5 \tP)\epsilon\,,
\end{equation}
and the scalar equations of motion (\ref{speom1}) become
\begin{eqnarray}
&&-\ft23\sigma S + \lambda + \ft12 \beta \Big[ \ft23 S R + 4 \Box S
+ \ft43 (2\tA^\mu \partial_\mu \tP + \tP \nabla^\mu \tA_\mu) + \ft49 S
(2(S^2-\tP^2) - \tA^2)\Big)=0\,,\cr 
&&-\ft23\sigma \tP + \ft12 \beta \Big[ \ft23 \tP R + 4 \Box \tP + \ft43
(2\tA^\mu \partial_\mu S + S \nabla^\mu \tA_\mu) + \ft49 \tP (2(S^2-\tP^2) -
\tA^2)\Big)=0\,,\label{speom2}
\end{eqnarray}

  Since many of the solutions that we shall obtain arise, with minor
differences as noted, both in the $U(1)$ supergravity theory and in the
$O(1,1)$ fake supergravity theory, we shall sometimes use the generic
term ``supersymmetric'' for both cases.  It should always be understood that
in the case of the $O(1,1)$ theory the solutions are actually
pseudo-supersymmetric rather than truly supersymmetric.

\subsection{AdS$_4$ vacua}

There may exist several AdS$_4$ vacua in which $A_\mu=0$ and ${\cal M}$ is a constant. For the $U(1)$ theory,  the scalar equations of motion imply
\begin{equation}
(\lambda-\ft23\sigma {\cal M}) + \ft43 \beta {\cal M} (\Lambda +
\ft13 {\cal M}\bar {\cal M})=0\,,
\end{equation}
and so the constant ${\cal M}$ must be real, {\it i.e.} $P=0$.  The equation
is a cubic polynomial in $S$,  and so it has at least one real
solution, with the possibility of three real solutions. As we shall
see later, the supersymmetric AdS$_4$ vacuum has $\Lambda=-\ft13 S^2$.

The AdS$_4$ solution with $P=0=A_\mu$ is also a solution in the $O(1,1)$
theory.  In that theory, however, there exists also a vacuum
solution in which $\tP$ is non-vanishing, provided that $\lambda=0$.
A supersymmetric AdS$_4$ can also arise in this case, which we shall
discuss in the next section.

\section{Supersymmetric domain walls (membranes)}

In this section, we construct supersymmetric domain wall solutions.
The ansatz is given by
\begin{eqnarray}
ds^2 &=& dr^2 + a(r)^2 dx^\mu dx_\mu\,,\cr 
A&=&\phi(r) dr\,,\qquad S=S(r)\,,\qquad P=P(r)\,.
\end{eqnarray}
Note that if $A_\mu$ were a massless gauge field, it would be
pure gauge.  However, since $A_\mu$ is a massive field, the
ansatz is nontrivial. A natural choice for the vielbein is $e^{\bar
r}=dr$, $e^{\bar \mu} = a dx^\mu$. The only non-vanishing components of
the corresponding spin connection
are then given by $\omega^{\bar
\mu}{}_{\bar r} = (a'/a) e^{\bar \mu}$, where a prime denotes a
derivative with respect to $r$.  For the $U(1)$ theory, the Killing
spinor equations become
\begin{eqnarray}
&&\partial_r\epsilon +\ft{\rm i}3\phi \epsilon + \ft16 \Gamma_r (S + {\rm i}
\Gamma_5 P)\epsilon=0\,,\cr 
&&\partial_\mu\epsilon + (\ft12 a' - \ft{\rm i}{6} a\phi) \Gamma_{\bar
\mu\bar r}\epsilon + \ft16 a \Gamma_{\bar \mu} (S + {\rm i}\Gamma_5 P)
\epsilon=0\,.
\end{eqnarray}

\subsection{Domain wall with a scalar potential}

Let us first consider $\phi=0=P$, which applies for
both the $U(1)$ and $O(1,1)$ theories. The solution is
supersymmetric provided that
\begin{equation}
S=\fft{3a'}{a}\,.
\end{equation}
The corresponding Killing spinor is subject to the projection
\begin{equation}
(\Gamma_{\bar r} + 1)\epsilon=0\,.
\end{equation}
We find that all the equations of motion then reduce to
\begin{equation}
\lambda a^2 -2 \sigma a a' + 6 \beta (a a''' - a' a'')=0\,.
\label{dwsusyeom}
\end{equation}
If $\sigma=0=\lambda$, then the equations of motion reduce simply to
\begin{equation}
a a''' - a' a'' =0\,,
\end{equation}
for which the general solution is given by
\begin{equation}
a= a_1 \cosh k r + a_2 \sinh k r \,,\qquad \hbox{or}\qquad a=\tilde
a_1 \cos k r + \tilde a_2 \sin k r\,.
\end{equation}
The second choice gives a solution with a naked power-law singularity and
we shall not consider it further.  For the first choice,
we find that not only do AdS$_4$ vacua with an arbitrary cosmological
constant arise, but AdS$_4$ wormholes can arise also.

If both $\sigma\ne 0$ and  $\lambda\ne0$, then the vacuum solution is
AdS$_4$ with $a=\exp(\lambda/(2\sigma))$. If $\lambda=0$, but $\sigma\ne 0$, the vacuum solution is
Minkowski spacetime with $a$ being constant.  If $\sigma=0$ and
$\lambda\ne 0$, neither AdS$_4$ nor Minkowski spacetime is a solution.

\subsection{Domain wall with $A_\mu\ne 0$}

We now consider the case with non-vanishing $\phi$ and $P$.  The
existence of a Killing spinor implies
\begin{eqnarray}
\hbox{$U(1)$ theory}:&& \ft12 (a' - \ft{\rm i}3 a \phi)^2 - \ft1{18}
(S^2 + P^2) =0\,,\cr 
\hbox{$O(1,1)$ theory}:&&\ft12 (a' + \ft13 a \tilde\phi)^2 - \ft1{18} (S^2
- \tP^2) =0\,,
\end{eqnarray}
where in the $O(1,1)$ theory the ansatz for the vector field becomes
$\tA = \tilde\phi\, dr$.
We see that for the $U(1)$ theory, the solution cannot be real
if $\phi$ and $P$ are non-vanishing.  This reality problem is resolved
in the $O(1,1)$ theory.  Substituting the supersymmetry condition
into the bosonic equations of motion, we find that if we set $S=\tP$,
the equations are reduced to
\begin{equation}
\sigma a a' - 3\beta (5a' a'' + a a''')=0\,.\label{dweom2}
\end{equation}
Note that as mentioned in section 2, turning on $\tP$ means we must have
$\lambda=0$.  The function $S=\tP$ is determined by
\begin{equation}
\sigma a S - 3 \beta (5a' S' + a S'')=0\,.
\end{equation}
Note that there is no back reaction of the scalars on
the metric, and hence the domain wall is supported by the vector
field alone.

    It is clear from (\ref{dweom2}) that Minkowski spacetime is a
vacuum solution.  It also admits an AdS solution with $a=e^{k r}$, where
\begin{equation}
k^2 = \fft{\sigma}{18\beta}\,.
\end{equation}

\section{Lifshitz solutions, and their
(pseudo)-supersymmetry}

In this section we study Lifshitz solutions following from the ansatz
\begin{eqnarray}
ds^2&=& \ell^2 \Big( \fft{dr^2}{r^2} - r^{2z} dt^2 + r^2 (dx^2 +
dy^2)\Big)\,,\cr 
A&=& q r^{z} dt + p \fft{dr}{r}\,,
\end{eqnarray}
in the $U(1)$ theory,
where $p$, $q$ and the scalars $S$ and $P$ are constants.  Note that
since $A_\mu$ is massive, with no gauge symmetry,
the $p$ term is nontrivial even though it is exact.  In the $O(1,1)$ theory
the ansatz for the vector field becomes
\be
\tA= \tilde q r^z dt + \tilde p \fft{dr}{r}\,,\label{o11case}
\ee
where $A$ is now written as $A={\rm i} \tA$ with $\tA$, and hence $\tilde q$
and $\tilde p$, being real.

    Lifshitz solutions were proposed in \cite{kacliumul} as gravity duals for
non-relativistic field theories. (See also \cite{korlib}.)  Although
Lifshitz solutions can be embedded in string theories and supergravities
\cite{Hartnoll}-\cite{Amado:2011nd}, supersymmetric Lifshitz solutions
are rare.  Lifshitz solutions arise naturally in higher-derivative
gravities.  It was shown in \cite{lppp} that not only the homogeneous 
Lifshitz vacua, but
also asymptotically Lifshitz black holes, can arise in Einstein-Weyl gravity.

\subsection{List of solutions}

\subsubsection{Solutions with $A_\mu=0$}

There are two classes of solutions with $A_\mu=0$.  The first is
when $P=0$, for which
\begin{eqnarray}
\lambda&=&\fft{2S}{9\ell^2} \Big(2\alpha z(z-4) + 3\beta
\Big(3(z^2+2z+3)-\ell^2 S^2 \Big)\Big)\,,\cr \sigma
&=&\fft{1}{3\ell^2}\Big( 2\alpha z(z-4) + \beta
\Big(6(z^2+2z+3)-\ell^2 S^2 \Big)\Big)\,,\cr S^2&=&\fft{3(z^2 + 2z +
3)}{2\ell^2}\,,\qquad \hbox{or}\qquad S^2 = \fft{\alpha z(z-4) +
3\beta(z^2+2z+3)}{2\beta\ell^2}\,.\label{A=0P=0}
\end{eqnarray}
The second class is when $P\ne 0$.  This implies that we must have
$\lambda=0$.  There are then two solutions:
\begin{eqnarray}
\sigma &=& \fft{\alpha z(z-4)}{3\ell^2}\,,\qquad \tP^2 -S^2=
\fft{3(z^2 + 2z + 3)}{\ell^2}\,, \qquad \beta=-\fft{\alpha
z(z-4)}{9 (z^2 + 2z + 3)}\,;\cr 
&&\hbox{or} \cr 
\sigma&=&0\,,\qquad S^2 + P^2 = \fft{3(z^2 + 2z +
3)}{2\ell^2}\,,\qquad \beta=-\fft{4\alpha z(z-4)}{9 (z^2 + 2z +
3)}\,.
\end{eqnarray}
Note that the first solution arises only for the $O(1,1)$ theory.  The
second solution, which is presented for the $U(1)$ theory, can also be
a solution in the $O(1,1)$ theory provided that $S^2+P^2$, which then becomes
$S^2-\tP^2$, is non-negative.
For $z(z-4)=0$, we
must have $\beta=0$.  This implies that $P=0$, and hence the
solution reduces to a special case of (\ref{A=0P=0}).

Next, we shall consider solutions with non-vanishing $A_\mu$.  We
find that the reality of the solution typically tends to
select the $O(1,1)$ rather than the $U(1)$ theory.

\subsubsection{$A_\mu\ne 0$ and $\alpha\ne 0$}

For non-vanishing $\alpha$, we find that the equations of motion
imply that either $p=0$ or $q=0$.  We then find solutions as follows:
First, we can take $p=0=P$, with $q$ non-vanishing.  We find
\begin{eqnarray}
&&\tilde q = z-1\,,\quad S=\fft{z+2}{\ell}\,,\quad  \sigma
\ell^2=\beta(z+2)^2 - 2\alpha z\,,\quad \lambda = \fft{2\sigma
(z+2)}{3\ell}\,;\label{p0P0sol1}\\ 
&&\tilde q=z-1\,,\quad \sigma=\beta S^2\,,\quad \lambda =\fft{2\beta
S(\ell^2 S^2 + 2 (z+2)^2)}{9\ell^2}\,,\cr 
&&\qquad\qquad\qquad 3\alpha z + 2\beta \Big((z+2)^2 - \ell^2
S^2\Big)=0\,;\label{p0P0sol2}\\ 
&& \tilde q^2= \Big(3(z^2+2z+3) - 2\ell^2 S^2\Big)\,,\quad \sigma =
\beta S^2\,,\quad \lambda = \ft23 \beta S^3\,,\quad z\alpha =0\,.
\label{p0P0sol3}
\end{eqnarray}
Note that of the above three solutions, the first two are for the
$O(1,1)$ theory, with the ansatz for the vector now taking the form
(\ref{o11case}).  The third solution,
with $\alpha=0$, which is presented for
the $O(1,1)$ theory, could also be real in the  $U(1)$ theory if the
right-hand side of the expression for $\tilde q^2$ were negative.

If instead
$q=0=P$, we find that there is a solution in the $O(1,1)$ theory,
given by
\begin{equation}
\tilde p=9\,,\quad z=4\,,\quad \sigma = \fft{108\beta}{\ell^2}\,,\quad
\lambda=0\,,\quad S=0\,.\label{q0P0sol1}
\end{equation}

Now we consider the case with non-vanishing $P$.  For this,
we find that the equations of motion always require that $\lambda=0$.
For $p=0$, we find two solutions in the $O(1,1)$
theory:
\begin{eqnarray}
&&\tilde q=z-1\,,\quad S^2 - \tP^2 = \fft{(z+2)^2}{\ell^2}\,,\quad
\sigma=0\,,\quad 2\alpha z = \beta (z+2)^2\,;\label{p0Pn0sol1}\\
&&\tilde q=z-1\,,\quad S^2 - \tP^2 = -\fft{2(z+2)^2}{\ell^2}\,,\quad \sigma
\ell^2 = -2 \beta (z+2)^2\,,\quad \alpha z =2 \beta
(z+2)^2\,,\label{p0Pn0sol2}
\end{eqnarray}
For $\tilde q=0$, we find a solution in the $O(1,1)$ theory, given by
\begin{eqnarray}
&& z=4\,,\qquad \sigma= \fft{\beta (\tilde p^2+6\tilde p + 81)}{2\ell^2}\,,
\qquad
\lambda = \fft{2\beta \tilde p\, (\tilde p-9)^2}{\tP\ell^4} \,,\nn\\
&&S=\fft{(\tilde p+9)^2}{12\tilde p}\, \tP\,,\qquad \tP=\pm
  \fft{6\sqrt2 \,\tilde p(\tilde p-9)}{
               \sqrt{(\tilde p+3)(\tilde p+27)(\tilde p^2+6\tilde p+81)}}\,.
\end{eqnarray}

\subsubsection{$A_\mu\ne 0$ and $\alpha= 0$}

In this case, we find solutions in the $O(1,1)$ theory with both $\tilde p$
and $\tilde q$ non-vanishing, given by
\begin{eqnarray}
&&\sigma = \fft{\beta \tilde p (z+2) (S\pm \tP)}{\ell^2 \tP}\,,\qquad \lambda =
\fft{2\beta \tilde p (z+2)(S^2-\tP^2)}{3\ell^2 \tP}\,,\cr 
&&\tilde q^2-\tilde p^2 = 3(z^2+2z+3) -
    \fft{\tilde p(z+2)(2S \pm 3\tP)}{\tP}\,,\cr 
&& S^2-\tP^2 - \fft{\tilde p(z+2)(S\pm 3\tP)}{\ell^2 \tP}=0\,.
\end{eqnarray}
A special case arises if $S=\tP$, implying $\lambda=\sigma=0$ and $\tilde q^2
= 3(z^2+2z+3)$.

\subsection{(Pseudo-)supersymmetry analysis}

Having obtained a variety of Lifshitz solutions in quadratic
curvature supergravity, we now examine their (pseudo-)supersymmetry.
Since they arise
mostly in the $O(1,1)$ theory, we shall present the analysis within this
framework. For simplicity, and without loss of generality,
we shall set $\ell=1$. A natural choice for
the vielbein is given by
\begin{equation}
e^{\hat 0}=r^z dt\,,\qquad e^{\hat x} = r dx\,,\qquad e^{\hat y} =r
dy\,,\qquad e^{\hat r}=\fft{dr}{r}\,,
\end{equation}
where we use hats to denote tangent space indices.
The non-vanishing components of the corresponding torsion-free spin connection
are then given by
\begin{equation}
\omega^{\hat 0}{}_{\hat r}= z e^{\hat 0}\,,\qquad \omega^{\hat
x}{}_{\hat r}= e^{\hat x}\,,\qquad \omega^{\hat y}{}_{\hat r}=
e^{\hat y}\,.
\end{equation}
The Killing spinor equations are
\begin{eqnarray}
&&\partial_t \epsilon + \ft12 zr^z \Gamma_{0\hat r} -\ft16 r^z (2\tilde q -
\tilde p \Gamma_{0\hat r}) \Gamma_5 \epsilon + \ft16 r^z \Gamma_{\hat 0} (S
+ \tP \Gamma_5)=0\,,\cr 
&&\partial_i \epsilon +\ft12 r \Gamma_{\hat i\hat r} \epsilon +
\ft16 r (\tilde p \Gamma_{\hat i \hat r} - \tilde q \Gamma_{\hat i \hat
0})\Gamma_5 \epsilon + \ft16 r \Gamma_{\hat i} (S + \tP \Gamma_5)
\epsilon=0\,,\cr 
&&\partial_r \epsilon - \ft16 r^{-1} (2\tilde p + \tilde q \Gamma_{\hat r\hat
0})\Gamma_5 \epsilon + \ft16 r^{-1} (S + \tP \Gamma_5) \epsilon=0\,,
\end{eqnarray}
where $i=x,y$.

   To establish the supersymmetry of a solution, one need only
demonstrate the existence of a Killing spinor,
without necessarily solving for it explicitly.  This can be done by
examining the integrability conditions. We find that
\begin{eqnarray}
&&0=[\partial_x,\partial_y]\epsilon=\Gamma_{xy}
U_{xy}\epsilon\,,\qquad
0=[\partial_r,\partial_i]\epsilon=\Gamma_{ri} U_{ri}\epsilon\,,\cr
&&0=[\partial_t,\partial_i]\epsilon=r^{z+1} \Gamma_{0i}
U_{ti}\epsilon\,,\qquad
0=[\partial_r,\partial_t]\epsilon=r^{z-1}\Gamma_{r0}
U_{rt}\epsilon\,,
\end{eqnarray}
where
\begin{eqnarray}
U_{xy} &=& \ft1{18} r^2 (9 + \tilde p^2-\tilde q^2 -S^2 + \tP^2)
   + \ft19 r^2 \Big(3
\tilde p+ (\tilde q\Gamma_0 - \tilde p \Gamma_r) (S + \tP \Gamma_5)\Big)
   \Gamma_5\,,\cr 
U_{ri} &=& \ft1{18}(9-\tilde q^2 -S^2 + \tP^2) +
     \ft16 \tilde p \Gamma_5 -\ft{1}{18}
\tilde p \tilde q \Gamma_{0r} + \ft19 (\tilde q\Gamma_0 + \tilde p \Gamma_r)
  (S + \tP \Gamma_5)\Gamma_5\,,\cr 
U_{ti} &=&\ft{1}{18}(9z + \tilde p^2 - S^2 + \tP^2) -
\ft19 (\tilde q\Gamma_0 + \tilde p
\Gamma_r)(S + \tP \Gamma_5)\Gamma_5 \cr 
&&+ \ft16 \Big((z+1) \tilde p + z \tilde q \Gamma_{0r}\Big)
  \Gamma_5 - \ft1{18} \tilde p
\tilde q \Gamma_{0r}\,,\cr 
U_{rt}&=& \ft{1}{18}(9z^2 - S^2 + \tP^2) - \ft1{9} (\tilde q\Gamma_0 - \tilde p
\Gamma_r)(S + \tP \Gamma_5 )\Gamma_5 -
  \ft{z}{6} (\tilde p + 2\tilde q \Gamma_{0r})
\Gamma_5\,.
\end{eqnarray}

     It is now
straightforward to verify whether the Lifshitz solutions we have
obtained are (pseudo-)supersymmetric or not.
 For the $A_\mu=0$ solutions, we find
from the integrability conditions that the only supersymmetric
solution is the maximally-supersymmetric AdS$_4$ vacuum.  In what
follows, we shall enumerate the supersymmetric solutions with
non-vanishing $A_\mu$.

Let us first consider $\tilde p=0$ and $\tP=0$. We find that the
(pseudo-)supersymmetric solutions in general satisfy
\begin{equation}
\tilde q=z-1\,,\qquad S=z+2\,.
\end{equation}
The Killing spinor satisfies the projections
\begin{equation}
\Gamma_0 \Gamma_5\epsilon -\epsilon=0\,,\qquad \Gamma_r \epsilon +
\epsilon=0\,,\label{lifssusyproj1}
\end{equation}
and so in general the solution preserves $\ft14$ of the
(pseudo-)supersymmetry.  It is clear that such a Lifshitz solution does
exist, given by (\ref{p0P0sol1}).  There are two cases where a
supersymmetry enhancement occurs. We find that when $z=-2$ or $z=0$, the
fraction of preserved supersymmetry is doubled to $\ft12$, with now only
the single projection given by
\begin{eqnarray}
z=-2:&&\Gamma^0 \Gamma_5\epsilon-\Gamma_r \epsilon=0\,,\cr 
z=0:&& \Gamma^0 \Gamma_5\epsilon +\epsilon=0\,.
\end{eqnarray}

   Interestingly, there is a maximally-(pseudo-)supersymmetric solution that is
not AdS$_4$.  It is given by (\ref{p0P0sol3}) with $z=0$ and $\tilde q=-3$,
and hence $S=0$.  Thus we have $\sigma=0$, and so the theory itself is
constructed from only quadratic super-invariants.
The four Killing spinors can be solved explicitly, and are given by
\begin{eqnarray}
\epsilon &=& \Big(1 - \ft12 r (x\Gamma_x + y\Gamma_y)(\Gamma_r +
\Gamma_0\Gamma_5)\Big)\eta\,,\cr 
\eta &=& \fft{e^t}{\sqrt{r}} \eta^+_1 + e^{-t}\sqrt{r} \,\eta^-_1 +
e^t\sqrt{r}\, \eta^+_2 + \fft{e^{-t}}{\sqrt{r}} \eta^-_2\,,
\end{eqnarray}
where $\eta^\pm_i$ are four constant spinors satisfying
\begin{equation}
(\Gamma_5 \pm 1) \eta^\pm _i = 0\,,\qquad (\Gamma_{01} -1)
\eta_1^\pm=0\,,\qquad (\Gamma_{01} + 1)\eta_2^\pm=0\,.
\end{equation}

Finally, we find that the solution (\ref{p0Pn0sol1}) also preserves
$\ft14$ of the (pseudo-)supersymmetry.  The Killing spinors are subject to
the constraints
\begin{equation}
\Big(S + \tP \Gamma_5  + (z+2) \Gamma_r\Big)\epsilon=0\,,\qquad
(\Gamma_0\Gamma_5 + \Gamma_r)\epsilon=0\,.\label{lifssusyproj4}
\end{equation}
It is clear that this projection reduces to (\ref{lifssusyproj1})
when $\tP=0$.  However, we nevertheless treat these as two separate
classes of solutions since turning on $\tP$ will force $\lambda=0$ in
the bosonic equations of motion.

Thus we have obtained all the (pseudo-)supersymmetric Lifshitz solutions in the
off-shell ${\cal N}=1$ supergravities that use
both the quadratic super-invariants.  The (pseudo-)supersymmetric Lifshitz solutions in Einstein-Weyl supergravity was obtained in \cite{luwangsusylif}.

\section{(Pseudo-)supersymmetric $T^2$-symmetric solutions}

In this section, we construct pseudo-supersymmetric $T^2$-symmetric
solutions in the O(1,1) theory. The ansatz is given by
\begin{equation}
ds^2 = \fft{dr^2}{f^2} - a^2 dt^2 + r^2 (dx^2 + dy^2)\,,\qquad
\tA= \phi dt\,.
\end{equation}
This ansatz encompasses all the Lifshitz solutions we obtained in
the previous section that have $\tilde p=0$.  We shall not include a term
$\tilde\psi(r) dr$ in the ansatz for the vector field $\tA_\mu$ here,
because in this section we shall
concentrate only on the pseudo-supersymmetric $T^2$-symmetric solutions.
As we have seen in the previous section, there is no (pseudo-)supersymmetric
Lifshitz solution that has non-vanishing $\tilde p$.

\subsection{(Pseudo-)supersymmetry conditions and equations of motion}

As in \cite{luwangsusylif}, the vielbein and the corresponding spin connection are given by
\begin{eqnarray}
&& e^{\hat r} = f^{-1} dr\,,\qquad e^{\hat 0} = a dt\,,\qquad
e^{\hat x}=r dx\,,\qquad e^{\hat y} = r dy\,,\cr 
&& \omega^{\hat 0}{}_{\hat r} = \fft{a'f}{a}\, e^{\hat 0}\,,\qquad
\omega^{\hat i}{}_{\hat r} =\fft{f}{r}\, e^{\hat i}\,,\qquad
(i=x,y)\,,
\end{eqnarray}
where a prime denotes a derivative with respect to $r$.  The Killing
spinor equations are given by
\begin{eqnarray}
\Big(\partial_t + \ft12 a' f \Gamma_{\hat 0\hat r} - \ft13\phi
\Gamma_5 + \ft16a\Gamma_{\hat 0} (S + \tP \Gamma_5)\Big)\epsilon
&=&0\,,\cr 
\Big(\partial_r + \fft{\phi}{6af} \Gamma_{\hat 0\hat r}\Gamma_5 +
\fft{1}{6f} \Gamma_r (S + \tP \Gamma_5)\Big)\epsilon &=& 0\,,\cr 
\Big(\partial_i + \ft12f\Gamma_{\hat i\hat r} + \fft{r\phi}{6a}
\Gamma_{\hat 0\hat i} \Gamma_5+ \ft16 r\Gamma_i (S +
\tP \Gamma_5)\Big)\epsilon &=&0\,.
\end{eqnarray}
Following a similar strategy to the one we used for obtaining supersymmetric
Lifshitz solutions, we find that for $\tP=0$, the existence of a
Killing spinor implies
\begin{equation}
\tP=0:\qquad \phi = \fft{(ra' - a)f}{r}\,,\qquad \fft{a'}{a} =
\fft{S}{f} - \fft{2}{r}\,.
\end{equation}
The scalar equation then gives
\begin{equation}
3\lambda - 2\sigma S + 2\beta f (2S S' + 3 f' S' + 3 f S'')=0\,,
\end{equation}
and the vector equation of motion gives
\begin{eqnarray}
0&=&r \sigma (r S -3f) -\beta r (r S - 3 f) (2 f S' + S^2)\cr
&&-\alpha f \Big( r f (3f'' - r S'') - r (r f' + 2 r S - 3 f) S' +
(S-3f')(5f-2r S - r f')\Big)\,.
\end{eqnarray}
The Einstein equations of motion are then all satisfied.  The
Killing spinor is given by $\epsilon=\sqrt{r} \epsilon_0$, and it
satisfies the projection (\ref{lifssusyproj1}).

For $\tP\ne 0$, the existence of a Killing spinor implies that
\begin{eqnarray}
\tP\ne 0:&&\phi = \fft{(ra' - a)f}{r}\,,\qquad S^2 - \tP^2 =
\Big(\fft{a'}{a} + \fft{2}{r}\Big)^2 f^2\,.
\end{eqnarray}
The Killing spinor is again given by $\epsilon=\sqrt{r} \epsilon_0$, but now
satisfying the projections
\begin{equation}
\Big[S+ \tP \Gamma_5  + \Big(\fft{a'}{a} + \fft{2}{r}\Big)
f\Gamma_{\hat r}\Big]\epsilon=0\,,\qquad (\Gamma_0\Gamma_5 +
\Gamma_{\hat r})\epsilon=0\,.
\end{equation}
For non-vanishing $\tP$, we find, after imposing the supersymmetry
conditions, that we must have $\lambda=0$ and furthermore $\tP$ is
a constant multiple of $S$.  This may be parametrised as
\begin{equation}
\tP(r)=\sin\theta\, S(r)\,,\label{PasS}
\end{equation}
where $\theta$ is a constant.  The scalar and vector equations now become
\begin{eqnarray}
0&=& \sigma S - \beta (3f^2 S'' + 3f'S' + 2\cos\theta S S')\,,\cr
0&=&r \sigma (r S \cos\theta -3f) -\beta r \cos\theta (r S \cos\theta- 3 f)
(2 f S' + S^2 \cos^2\theta)\cr
&&-\alpha f \Big[ r f (3f'' - r S''\cos^2\theta) -
r (r f' + 2 r S \cos\theta - 3 f) S' \cos\theta\cr
&&\qquad\quad + (3f'-S\cos\theta)(r f' + 2r S \cos\theta -5f)\Big]\,.
\label{sfeom}
\end{eqnarray}
Note that when $\theta=0$ we have $\tP=0$, but the equations are reduced to
the previous $\tP=0$ case only for $\lambda=0$.

\subsection{Some exact solutions}

First, we consider the case where $\tP=0$.  Setting $\lambda=2$ and $\sigma=1$,
we obtain the solution
\begin{equation}
f=r-r_0\,,\qquad a=\fft{(r-r_0)^3}{r^2}\,,\qquad
\phi=\fft{3(r-r_0)^3 r_0}{r^3}\,,
\end{equation}
provided that $\beta = 1/9$. This is also a solution of conformal
supergravity with $\sigma=\lambda=\beta=0$ \cite{luwangsusylif}.

Now consider instead when $\tP\ne 0$.  In this case, $\tP$ is given
by (\ref{PasS}). One particularly simple situation
is when $\sin\theta=1$ and hence $\tP=S$.  The general solution for
the metric functions is then given by
\begin{equation}
a^2=\fft{1}{r^4}\,,\qquad f^2=c_0 + c_1 r^6 + \fft{\sigma r^2}{4\alpha}\,.
\end{equation}
It appears unlikely that the equations (\ref{sfeom}) are solvable exactly
in general, and we have not found any further exact solutions.

\section{Gyrating Schr\"odinger geometries}

In this section, we consider another class of homogeneous metrics,
namely the gyrating Schr\"odinger geometries  \cite{lupogyraton}.
The general ansatz is
\begin{eqnarray}
ds^2 &=&\ell^2 \Big[ \fft{dr^2 - 2 du dv + dx^2}{r^2} -
\fft{2c_2 du dx}{r^{z+1}} -\fft{c_1 du^2}{r^{2z}}\Big]\,,\cr
A&=&q \fft{dt}{r^z} + p \fft{dr}{r}\,,\label{schrans}
\end{eqnarray}
with the scalars $S$ and $P$ being constant.  In the case of the $O(1,1)$
theory, the ansatz for the vector $A$ will become $A={\rm i}\tA$ with
\be
\tA = \tilde q\, \fft{dt}{r^z} +\tilde p\, \fft{dr}{r}\,.\label{tAans}
\ee
The solution is
of Schr\"odinger type if $c_2=0$,
and the term $c_1$ adds a further deformation to the
Schr\"odinger metric. The metric
is AdS$_4$ if $z=1$.  There are two other Einstein metrics, given by
\begin{eqnarray}
z=-\ft12:&& c_2=0\,,\cr 
z=-2:&& c_1 + \ft12 c_2^2=0\,.
\end{eqnarray}
The first solution above is the Kaigorodov metric \cite{kaig}.
When $c_2=0$, the $z=2$
solution has Schr\"odinger symmetry and was proposed as a gravity dual for
the Schr\"odinger system \cite{Son:2008ye,Balasubramanian:2008dm}.  The
solutions of \cite{Son:2008ye,Balasubramanian:2008dm} make use of a
massive vector, which is absent in typical supergravities.  However,
a massive vector arises naturally in higher-order ${\cal N}=1$ $D=4$ off-shell
supergravity.  AdS gyratons were studied in
\cite{Frolov:2005ww}. Supersymmetric (gyrating) Schr\"odinger solutions
in Einstein-Weyl supergravity were constructed in \cite{lupogyraton,liulupp}.
Note that the metric of the gyrating Schr\"odinger solution 
(\ref{schrans}) is homogeneous, as is the Schr\"odinger metric.

We shall now present more general solutions that are not themselves
Einstein metrics.
As in the case of Lifshitz solutions, we shall present the bosonic
solutions first, and then study their supersymmetry.

\subsection{$A_\mu=0$}

In this subsection, we list solutions where the massive vector $A_\mu$ vanishes.
It can be easily verified that if $P\ne 0$, the scalar equations require
that $\lambda=0$.  Thus we shall consider first the case with $P=0$.
For the Schr\"odinger solutions (i.e with $c_2=0$), we then have
\begin{eqnarray}
S=\fft{3}{\ell}:&& \sigma=\fft{9\beta + 2\alpha z(1-2z)}{\ell^2}\,,\qquad
\lambda=\fft{18\beta + 4\alpha z(1-2z)}{\ell^3}\,,\cr
\sigma=\beta S^2:&& \lambda=\fft{2\beta S(18 + \ell^2 S^2)}{9\ell^2}\,,\qquad
2 \beta (\ell^2 S^2-9) + 3 \alpha z (2 z-1)=0\,.
\end{eqnarray}
For gyrating solutions, namely where $c_2\ne 0$, we find
\begin{eqnarray}
S=\fft{3}{\ell}:&& \sigma=\fft{9\beta -\alpha z(z+1)}{\ell^2}\,,
\quad \lambda=\fft{18\beta - 2\alpha z(z+1)}{\ell^3}\,,\cr
&& \alpha (2 c_1 + c_2^2)z(1+2z)=0\,;\\
\sigma=\beta S^2:&&c_1+\ft12c_2^2= 0\,,\qquad
 \lambda=\fft{2 \beta S (18 + \ell^2 S^2)}{9 \ell^2}\,,\cr
&&3\alpha z(z+1) + 4\beta (\ell^2 S^2-9)=0\,.
 \end{eqnarray}
Since $P$ and $A_\mu$ are both vanishing here, it follows that these solutions
arise in both the $U(1)$ and the $O(1,1)$ theories.

Now consider the case with $P\ne 0$, for which we must have $\lambda=0$.
We find two Schr\"odinger solutions:
\begin{eqnarray}
&&S^2 +P^2 = \fft{9}{\ell^2}\,, \qquad \beta=\ft29 \alpha z (2z-1)\,,
\qquad \sigma=0\,; \label{sch1}\\
&&\tP^2 -S^2 = \fft{18}{\ell^2}\,,\qquad \beta=\ft1{18} \alpha z (2z-1)\,,
\qquad \sigma=\fft{\alpha z(1-2z)}{\ell^2}\,.\label{sch2}
\end{eqnarray}
In addition, there are two types of gyrating solution:
\begin{eqnarray}
S^2 +P^2 = \fft{9}{\ell^2}:&& (2 c_1 + c_2^2)z(1+2z)=0\,,\quad
\beta=\ft19 \alpha z (z+1)\,,\quad \sigma=0\,;\label{gyr1}\\
\tP^2-S^2 = \fft{18}{\ell^2}:&&(2c_1+c_2^2)z(1+2z)=0\,,
\quad \beta=\ft1{36} \alpha z (z+1)\,,\cr
&& \sigma=-\fft{\alpha z(z+1)}{2\ell^2}\,.\label{gyr2}
\end{eqnarray}
The solutions (\ref{sch1}) and (\ref{gyr1}) are presented in the $U(1)$ theory,
but they could also arise in the $O(1,1)$ theory, with $P=-{\rm i} \tP$,
provided that $\tP^2$ is sufficiently small that $S^2-\tP^2$ remains
non-negative. The solutions (\ref{sch2}) and (\ref{gyr2}) can only arise in
the $O(1,1)$ theory.

\subsection{$A_\mu\ne 0$}

     When $A_\mu$ is turned on, as in the ansatz (\ref{schrans}), we find
that the equations of motion imply the constraints
\begin{equation}
(z+1)\, \alpha\, p\, q = 0\,.
\end{equation}
Solutions then arise as follows:

\bigskip
\noindent{\bf Case 1: $\alpha \ne 0$}:
\medskip

In this case, and if or $p=0=P$, we find
\begin{eqnarray}
S=\fft{3}{\ell}:&& q=\fft32 \sqrt{2c_1+c_2^2}\,(z-1)\,,\qquad
\sigma = \ft12\ell\lambda=\fft{9 \beta - \alpha z (z+1)}{\ell^2}\,;
\label{schrsusy1}\\
\sigma=\beta S^2:&&
\lambda =\fft{2 \beta S (18 + \ell^2 S^2)}{9 \ell^2}\,,\qquad
q=\ft32 \sqrt{2 c_1 + c_2^2}\, (z-1)\,,\cr
&&3\alpha z(z+1)+4\beta (\ell S^2 -9)=0\,.
\end{eqnarray}
For $p=0$, but $P\ne 0$, we must have $\lambda=0$.  The solutions are
\begin{eqnarray}
S^2 + P^2 = \fft{9}{\ell^2}: && q=3/2 \sqrt{2 c_1 + c_2^2}\, (z-1)
\,,\qquad\beta=\ft19 \alpha z(z+1)\,,\quad \sigma=0\,;\label{schrsusy2}\\
\tP^2 -S^2 = \fft{18}{\ell^2}: && q=3/2 \sqrt{2 c_1 + c_2^2}\, (z-1)\,,
\qquad\beta=\ft1{36} \alpha z(z+1)\,,\cr
&&\sigma=-\fft{\alpha z(z+1)}{2\ell^2}\,.
\end{eqnarray}
The first solution, written for the $U(1)$ theory, can arise also for the
$O(1,1)$ theory with $P=-{\rm i}\tP$ and $A_\mu={\rm i} \tA_\mu$, provided
that $S^2-\tP^2$ is still non-negative.  The second solution arises only in
the $O(1,1)$ theory.

If instead $q=0$ and $p\ne0$, we have
\begin{eqnarray}
\lambda=\fft{2\beta p (P^2 + S^2)}{\ell^2 p}\,,\qquad
\sigma=\fft{\beta (6 p S + \ell^2 P (P^2 + S^2))}{3 \ell^2 P}\,,\cr
(-18 + p^2) P + 3 p S + \ell^2 P (P^2 + S^2)=0\,,\cr
12\ell^2 p S (p^2 + S^2) - P (p^2 + 9) (p^2 +36)=0\,,\cr
c_1 = -\fft{c_2^2(z-1)(7z+4)}{4(2z+1)(2z-1)}\,,\qquad
z=0,\pm1,-2\,,\label{case1}
\end{eqnarray}
for the $U(1)$ theory.  In the $O(1,1)$ theory, we have
\begin{eqnarray}
&&\lambda = -\fft{\beta \tilde p(\tilde p-3)(\tilde p-6)}{\tP}\,,\qquad
\sigma=\ft12\beta (\tilde p^2+3\tilde p+15)\,,\qquad  z=0,\pm1,-2\,,\nn  \\
&&S=-\fft{(\tilde p^2+9\tilde p+18)\, \tP}{2\tilde p}\,,\qquad
  \tP^2= \fft{18 \tilde p^2\, (\tilde p-3)(\tilde p-6)}{
   (\tilde p^2+3\tilde p+18)(\tilde p^2+15\tilde p+18)}\,,
\end{eqnarray}
where now the ansatz for the vector field in (\ref{schrans}) is written in
terms of the tilded field $\tA_\mu$ as in (\ref{tAans}).
In the cases $z=-1$ and $z=-2$ there is a further constraint, namely
\begin{equation}
c_1 = -\fft{c_2^2(z-1)(7z+4)}{4(2z+1)(2z-1)}\,.
\end{equation}

\bigskip
\noindent{\bf Case 2: $\alpha = 0$}:
\medskip

In this case, we find that $pq$ can be non-zero, and the solution in
the $U(1)$ theory is given by
\begin{eqnarray}
\lambda=\fft{2\beta p (P^2 + S^2)}{\ell^2 P}\,,\qquad
\sigma=\fft{\beta \Big(6 p S + \ell^2 P (P^2 + S^2)\Big)}
{3 \ell^2 P}\,,\cr
18 P - p^2 P - 3 p S - \ell^2 P (P^2 + S^2)=0\,,\cr
2 (p^4-63p^2+81) P + 9 p (p^2-9) S +3\ell^2(P^2 + S^2)
(9 P + p S - \ell^2 P (P^2 + S^2))=0\,,\label{case2}
\end{eqnarray}
In the $O(1,1)$ theory, we have
\begin{eqnarray}
&&\lambda = -\fft{\beta \tilde p(\tilde p-3)(\tilde p-6)}{\tP}\,,\qquad
\sigma=\ft12\beta (\tilde p^2+3\tilde p+15)\,,\nn\\
&&S=-\fft{(\tilde p^2+9\tilde p+18)\, \tP}{2\tilde p}\,,\qquad
  \tP^2= \fft{18 \tilde p^2\, (\tilde p-3)(\tilde p-6)}{
  (\tilde p^2+3\tilde p+18)(\tilde p^2+15\tilde p+18)}\,.
\end{eqnarray}
It is of interest to note that there is no restriction on
the parameters $z, c_1$ and $c_2$ in either of these solutions.

\subsection{Supersymmetry analysis}

To examine the supersymmetry of the solutions we have obtained
in this section, we choose the vielbein
\begin{equation}
e^+ = du\,,\qquad e^- = \fft{dv}{r^2} + \fft{c_2 dx}{r^{z+1}}
+ \fft{c_1 du}{2r^{2z}}\,,\qquad e^{\hat r}=\fft{dr}{r}\,,\qquad
e^{\hat x}= \fft{dx}{r}\,,
\end{equation}
such that the metric is given by $ds^2=-2 e^+e^- + e^{\hat x} e^{\hat x} +
e^{\hat r} e^{\hat r}$. Note that for simplicity, we have set $\ell=1$.
 The corresponding spin connection has non-vanishing components given by
\begin{eqnarray}
&&\omega^{\hat x}{}_{\hat r} = -e^{\hat x} + \fft{(z-1)c_2}{2r^z} e^+\,,
\qquad \omega^{\hat x}{}_+=\fft{(z-1)c_2}{2r^z} e^{\hat r}\,,\qquad
\omega^{\hat r}{}_- = - e^+\,,\cr 
&&\omega^{\hat r}{}_+ = -\fft{(z-1)c_2}{2r^z} e^{\hat x} -\fft{(z-1)c_1}{r^{2z}}
e^+ - e^-\,,\qquad \omega^+{}_+=-e^{\hat r}\,,
\end{eqnarray}
and so the components of the Killing spinor equation are given by
\begin{eqnarray}
0&=&\partial_u \epsilon + \ft{(z-1) c_2}{4 r^{z}} \Gamma_{\hat x\hat r}
\epsilon +
\ft12 \Gamma_{\hat r+} \epsilon - \fft{(2z-1)c_1}{4r^{2z}} \Gamma_{-\hat r}
\epsilon
+ \ft16 (\Gamma_+ + \fft{c_1}{2r^{2z}}\Gamma_-)
(S + {\rm i} P\Gamma_5) \epsilon\cr 
&&+
\ft{\rm i}6 \Big(\fft{q}{r^z}(2 + \Gamma_{+-}) - p(\Gamma_{+\hat r}
 + \fft{c_1}{2r^{2z}}
\Gamma_{-\hat r})\Big)\Gamma_5 \epsilon\,,\cr 
0&=& \partial_v \epsilon - \fft{1}{2r^2} \Gamma_{-\hat r} \epsilon -
\fft{\rm i\,p}{6r^2} \Gamma_{-\hat r} \Gamma_5 \epsilon + \fft{1}{6r^2}
\Gamma_- (S + {\rm i} P\Gamma_5) \epsilon\,,\cr 
0&=& \partial_r \epsilon +\fft{(z-1)c_2}{4r^{z+1}} \Gamma_{-\hat x} \epsilon +
\fft{1}{2r} \Gamma_{+-} \epsilon - \fft{\rm i\,q}{6r^{z+1}}\Gamma_{-\hat r}
\Gamma_5 \epsilon + \fft{{\rm i}\,p}{3r}\Gamma_5 \epsilon +
\fft{1}{6r} \Gamma_{\hat r} (S + {\rm i} P\Gamma_5)\epsilon\,,\cr
0&=& \partial_x \epsilon - \fft{1}{2r} \Gamma_{\hat x\hat r}\epsilon
- \fft{(z+1) c_2}{4r^{z+1}}
\Gamma_{-\hat r} \epsilon  +\ft16(\fft1{r} \Gamma_{\hat x} +
\fft{c_2}{r^{z+1}} \Gamma_-) (S + {\rm i} P\Gamma_5)
\epsilon\cr 
&& - \ft{\rm i}{6}\Big(\fft{q}{r^{z+1}} \Gamma_{-\hat x}
   + p(\fft1{r}\Gamma_{\hat x\hat r} +
\fft{c_2}{r^{z+1}} \Gamma_{-\hat r})\Big) \Gamma_5\epsilon\,.
\label{gyrkseom}
\end{eqnarray}

Having obtained the Killing spinor equations, we can study the integrability
conditions to determine whether there exists a Killing spinor for a particular
background.  The Killing spinor equations (\ref{gyrkseom}) can be expressed as
\begin{equation}
\partial_u \epsilon = U \epsilon\,,\qquad
\partial_v \epsilon = V \epsilon\,,\qquad \partial_x \epsilon = X \epsilon\,,
\qquad \partial_r \epsilon = R \epsilon\,.
\end{equation}
This implies, for example,
\begin{equation}
\partial_v\partial_u \epsilon = \partial_v U \epsilon + UV\epsilon\,,\qquad
\partial_u\partial_v \epsilon = \partial_u V \epsilon + VU\epsilon\,,
\end{equation}
and so
we have the following derivative-independent equation on $\epsilon$:
\begin{equation}
(\partial_v U - \partial_u V + [U,V])\epsilon=0\,.
\end{equation}
There are in total six such equations, from the possible pairs taken from
$\{U,V,X,R\}$.  Examining these integrability conditions
for $A_\mu=0$, we find that only the Schr\"odinger
solutions ({\it i.e.}~with $c_2=0$)
can have Killing spinors. For these solutions, supersymmetry requires that
$S^2 + P^2 =9$, and the Killing spinors satisfy the projections
\begin{equation}
\Gamma_- \epsilon =0\,,\qquad
\Gamma_{\hat r}\epsilon=\ft13 (S + {\rm i} P \Gamma_5)\epsilon\,.
\end{equation}
Thus there is one Killing spinor, and it depends on $r$ only.

  An interesting situation arises for these Schr\"odinger solutions in
the special case
$P=0$; {\it i.e.}~if $S=3$. It turns out that the integrability conditions are
then satisfied if $\epsilon$ obeys just the single projection
\be
\Gamma_- \epsilon =0\,,
\ee
which would suggest that there should be two Killing spinors.  However,
one finds in this case that the Killing spinor equations (\ref{gyrkseom})
themselves can only be solved if the second projection condition
\be
\Gamma_{\hat r}\epsilon=\epsilon
\ee
is also satisfied, and so there is in fact only a single Killing spinor in
this special case too.  This is an example, not often encountered in
practice in supergravity examples, where the second-order integrability
conditions obtained by commuting pairs of Killing-spinor derivatives are not
sufficient to determine the existence of solutions.  In principle, one
might have to look at third-order integrability conditions or beyond.  (For
a discussion of this in the supergravity context, see \cite{vannwarn}.) Of
course, if one explicitly constructs the most general solution of the
Killing-spinor conditions themselves, it is not necessary to examine the
higher-order integrability conditions.  In practice, as in this example,
projection conditions that one learns from the usual second-order
integrability conditions, even if they are not providing the complete set
of projections, can be helpful when constructing the Killing spinors
explicitly.

For $A_\mu\ne 0$, we find that supersymmetry requires $p=0$.
There are two inequivalent solutions. The first is given by
\begin{equation}
q=-\ft32 c_2 (z-1)\,,\qquad c_1=0\,, \qquad S^2 + P^2 = 9
\end{equation}
In this case, there is only one (constant) Killing spinor, subject to
the projection
\begin{equation}
\Gamma_+ \epsilon =0\,,\qquad \Gamma_{\hat r}\epsilon=
\ft13 (S + {\rm i} P \Gamma_5)\epsilon\,.
\end{equation}
The second solution is given by
\begin{equation}
q=\ft12 c_2 (z-1)\,, \qquad S^2 + P^2 = 9\,.
\end{equation}
There is again only one (constant) Killing spinor, subject to
the projections
\begin{equation}
\Gamma_- \epsilon =0\,,\qquad \Gamma_{\hat r}\epsilon=
\ft13 (S + {\rm i} P \Gamma_5)\epsilon\,.
\end{equation}
Comparing with the solutions obtained in the previous subsection,
it is straightforward to see that the solutions (\ref{schrsusy1}) and
(\ref{schrsusy2}) can be made supersymmetric. Both BPS solutions
in Einstein-Weyl supergravity $(P=0)$ were obtained in \cite{lupogyraton}.

\section{More general gyrating solutions}

\subsection{A general class of solutions}

In general, we can consider the following most general gyraton metrics:
\begin{eqnarray}
ds^2 &=&\ell^2 \Big[ \fft{dr^2 - 2 du dv + dx^2}{r^2} -
2h(r,u,x) du dx -H(r,u,x) du^2\Big]\,,\cr
A&=&\phi(r,u,x) du + \psi(r,u,x) dr\,,\qquad
S=S(r,u,x)\,,\qquad P=P(r,u,x)\,.
\end{eqnarray}
These  become pp-waves when $h=0$.  Such pp-wave solutions in
critical gravity and more general higher-derivative gravities can be found in
\cite{Alishahiha:2011yb,Gullu:2011sj,Gurses:2012db}.

The general equations of motion are rather complicated to present.
There is no $u$-derivative in any of the equations, and so all
``constants of integration'' can trivially be taken to be functions of
$u$.  For simplicity of notation, the freedom to add such arbitrary $u$
dependence will be understood, but not explicitly indicated. A
further simplification can be achieved by considering cases where
$P=0$ and $S$ is a constant, in which case two possible choices arise:
\begin{equation}
S^2=\fft{9}{\ell^2}\,,\qquad \hbox{or}\qquad
S^2=\fft{\sigma}{\beta}\,.
\end{equation}
The equations become completely solvable if we then make the further
assumption that the functions are all independent of
$x$, leading to the ansatz:
\begin{eqnarray}
ds^2 &=&\ell^2 \Big[ \fft{dr^2 - 2 du dv + dx^2}{r^2} - 2h(r,u) du dx
  -H(r,u) du^2\Big]\,,\cr
A&=&\phi(r) dt\,,\qquad
S={\rm const.}\,,\qquad P=0\,.
\end{eqnarray}
We then find that the functions $\phi$, $h$ and $H$ are given by
\begin{eqnarray}
\phi&=&\fft{q_1}{r^z} + q_2 r^{z+1}\,,\qquad h=h_1r^{-z-1} +
h_2 r^z + h_3 r^{-2} + h_4 r\,,\cr
H&=&H_1 r^{-z-1} + H_2 r^z + H_3r^{-2} + H_4 r\cr
&&-\ft12 h_4^2 r^4 - h_1 h_4 r^{2 - z} - \ft12 h_1^2 r^{-2 z} -
 h_2 h_4 r^{3 + z} - \ft12 h_2^2 r^{2 z+2}\cr
 &&+\fft{2 q_1^2}{9 (z-1)^2r^{2z}} + \fft{2 q_2^2 r^{2 z+2}}{
9 (z+2)^2}\,.\label{gyrgenericz}
\end{eqnarray}
For the constants, there are two possibilities:
\begin{equation}
\ell^2S^2 = 9\,,\qquad \sigma =\ft12\lambda\ell=
\fft{9 \beta -\alpha z ( z+1)}{\ell^2}\,,
\end{equation}
or
\begin{equation}
\sigma = \beta S^2\,,\qquad
\lambda = \fft{2\beta S (18 + \ell^2 S^2)}{9 \ell^2}\,,\qquad
3\alpha z(z+1) + 4\beta (\ell^2 S^2-9)=0\,.
\end{equation}
There exist two critical values of $z$, namely $z=1$ or $z=-2$, for which the
solution degenerates.  The functions $\phi$, $h$ and $H$ are now given
by
\begin{eqnarray}
\phi&=&\fft{q_1}{r} + q_2 r^{2}\,,\qquad
h=\fft{(h_1 + h_2r^3)\log r}{r^2} + \fft{h_3 + h_4 r^3}{r^2}\,,\cr
H&=& \fft{(H_1 + H_2r^3)\log r}{r^2} + \fft{H_3 + H_4 r^3}{r^2}
+\fft{2 (6 q_1^2 + q_2^2 r^6)}{81 r^2} +\fft{2 q_1^2 \log r
(4 + 3 \log r )}{27 r^2}\cr
&&-\fft{2 h_1^2 + 4 h_1 h_2 r^3 + 3 h_4^2 r^6}{6 r^2} -
\fft{(2 h_1^2 - 4 h_1 h_2 r^3 + 3 h_2 h_4 r^6) \log r}{3 r^2}\cr
&&-\fft{(h_1 + h_2 r^3)^2 (\log r)^2}{2 r^2}\,.\label{gyrcritical}
\end{eqnarray}

Logarithmic behavior can also arise when $z=-1/2$, for which we
have
\begin{eqnarray}
\phi &=& \sqrt{r} (q_1 + q_2 \log r)\,,\qquad h= \fft{(h_1 + h_2\log
r)}{\sqrt r} + \fft{h_3}{r^2} + h_4 r\,,\cr 
H &=& \fft{(H_1 + H_2\log
r)}{\sqrt r} + \fft{H_3}{r^2} + H_4 r +
\ft{16}{243} q_1 q_2 r (3\log r -5) + \ft{8}{729} q_2^2 r (3\log
r-2)^2\cr 
&& -\ft{1}{18} r \Big(-30 h_1 h_2 - 16 h_2^2 + 18 h_1 h_4 r^{3/2} + 9 h_4^2 r^3
\cr 
&&\qquad\qquad  +
   18 h_2 (h_1 + h_4 r^{3/2}) \log r + 9 h_2^2 (\log r) ^2\Big)\,.
\label{zmhalf}
\end{eqnarray}
Note that for these solutions, the parameters $h_3$ and $H_3$ are trivial.


\subsection{Supersymmetry analysis}

We shall choose the vielbein basis
\begin{equation}
e^+=du\,,\qquad e^- = \fft{dv}{r^2} +\ft12 H du + h dx \,,\qquad
e^{\hat r} = \fft{dr}{r}\,,\qquad e^{\hat x} = \fft{dx}{r}\,.
\end{equation}
The non-vanishing components of the spin connection are then given by
\begin{eqnarray}
&&\omega^{\hat x}{}_{\hat r} = - e^{\hat x} - (rh + \ft12 r^2 h')
e^+\,,\qquad
\omega^{\hat x}{}_+ = - (rh + \ft12 r^2 h') e^{\hat r}\,,\cr 
&&\omega^{\hat r}_+ = (r h + \ft12 r^2 h') e^{\hat x} +
(H + \ft12 r H') e^+ - e^-\,,\quad \omega^{\hat r}{}_- =
-e^{+}\,,\quad \omega^+{}_+ = - e^{\hat r}\,,
\end{eqnarray}
The Killing spinor equations are
\begin{eqnarray}
0&=& \partial_u \epsilon- \ft14(2rh + r^2 h') \Gamma_{\hat x\hat r} \epsilon
+\ft12 \Gamma_{\hat r +} \epsilon -
\ft14 (H + r H') \Gamma_{\hat r -}\epsilon +
\ft{\rm i}{6} \phi (\Gamma_{+-} + 2) \Gamma_5 \epsilon\cr 
&& +
\ft16 (\Gamma_+ + \ft12 H \Gamma_-)(S + {\rm i}
P\Gamma_5)\epsilon\,,\cr 
0&=& \partial_v \epsilon + \ft{1}{2r^2} \Gamma_{\hat r-} \epsilon +
\ft{1}{6r^2}\Gamma_- (S + {\rm i} P\Gamma_5)\epsilon\,,\cr 
0&=&\partial_r \epsilon + \ft14 (2h + r h')\Gamma_{\hat x -}
\epsilon + \ft1{2r} \Gamma_{+-} \epsilon + \ft{\rm i}{6r}\phi
\Gamma_{\hat r -}\Gamma_5 \epsilon +\ft{1}{6r} \Gamma_{\hat r}
(S + {\rm i} P \Gamma_5)\epsilon\,,\cr 
0&=&\partial_x\epsilon - \ft{1}{2r} \Gamma_{\hat x\hat r} \epsilon -
\ft14 r h' \Gamma_{\hat r -} \epsilon + \ft{\rm i}{6r} \phi
\Gamma_{\hat x -} \Gamma_5 \epsilon +
\ft16 (\ft{1}{r} \Gamma_{\hat x} + h \Gamma_-)(S + {\rm i} P
\Gamma_5)\epsilon\,.
\end{eqnarray}
For the solutions with $P=0$ we considered earlier,
it is clear that if we turn off $h$ and $\phi$, they then
reduce to a special class of AdS pp-waves and hence preserve $\ft14$ of the
supersymmetry, provided that $S=3/\ell$.  In fact, it was shown
in \cite{kerimo} that the most general pp-wave with $r$, $u$ and $x$ dependence and
with $A_\mu$ turned off all preserve $\ft14$ of the supersymmetry. The Killing
spinor satisfies the projections
\begin{equation}
\Gamma_{\hat r} \epsilon = \epsilon\,,\qquad
\Gamma_-\epsilon =0\,.\label{genschrks1}
\end{equation}
For non-vanishing $\phi$ and $h$, Killing spinors with the same
projections (\ref{genschrks1}) also exist, provided that
\begin{equation}
\phi = - \ft12 (2r h + r^2 h')\,.
\end{equation}
Thus the bosonic solution (\ref{gyrgenericz}) becomes
supersymmetric provided that the condition
\begin{equation}
\ft32 h_4 r^2 + r^{-z} \Big(q_1 - \ft12 (z-1)h_1\Big) + r^{z+1}
\Big(q_2 + \ft12(z+2)h_2\Big)=0
\end{equation}
holds for all $r$.  For generic $z$, we must therefore have
\begin{equation}
h_4=0\,,\qquad q_1=\ft12 (z-1) h_1\,,\qquad q_2=-\ft12 (z+2) h_2\,.
\end{equation}

For the critical solution (\ref{gyrcritical}), we find that
supersymmetry implies
\begin{equation}
q_1=-\ft12 h_1\,,\qquad h_2=0\,,\qquad q_2=-\ft32 h_4\,.
\end{equation}
For the $z=-1/2$ solution (\ref{zmhalf}), supersymmetry implies
\begin{equation}
h_4=0\,,\qquad q_1=-\ft14 (3h_1 + 2 h_2)\,,\qquad
q_2=-\ft34 h_2\,.
\end{equation}

Finally, we find that there exists another type of Killing spinor,
satisfying
\begin{equation}
\Gamma_{\hat r} \epsilon = \epsilon\,,\qquad
\Gamma_+\epsilon =0\,.\label{genschrks2}
\end{equation}
It requires that
\begin{equation}
\phi = -\ft32 r (2h + r h')\,,\qquad 2H + r H'=0\,.
\end{equation}
Applying this condition to the three solutions, we find that
$H_1=H_2=H_4=0$, and that
\begin{eqnarray}
\hbox{Generic $z$}:&& q_1=\ft32 (z-1) h_1\,,\quad q_2=-\ft32 (z+2)
h_2\,,\quad h_4=0\,,\cr
z=1,-2:&&q_1=-\ft32 h_1\,,\quad q_2=-\ft92 h_4\,,\quad
h_2=0\,,\cr
z=-\ft12:&& q_1=-\ft34 (3h_1 + 2 h_2)\,,\qquad q_2=-\ft94 h_2\,,\qquad
h_4=0\,.
\end{eqnarray}


\section{Conclusions}

In this paper, we have considered four-dimensional ${\cal N}=1$
off-shell supergravity including all four super-invariants
up to and including quadratic order in curvature.
These comprise a ``cosmological term,'' the Einstein-Hilbert term,
and two quadratic-curvature terms, one formed using the square of the
Weyl tensor, and the other formed using the square of the Ricci scalar.
In addition to the graviton and the
gravitino, the fields of the off-shell multiplet include a
complex scalar $S + {\rm i} P$
and a vector $A_\mu$.  In the Einstein plus cosmological supergravity, the
complex scalar and the vector are auxiliary and possess no physical degrees
of freedom.   The supersymmetric solution space is then rather limited.
Examples of such solutions were given in \cite{liuzayrei}.

    However, when the curvature-squared super-invariants are included,
the auxiliary fields can develop dynamics, and in particular the vector
becomes a massive Proca field.  For lack a more satisfactory name, one
may continue to call these fields auxiliary, even though they may now
propagate.  (The supersymmetry algebra still closes off-shell, however.)
In Einstein-Weyl supergravity,
Lifshitz solutions and also a new type of supersymmetric
gyrating Sch\"odinger
solution were obtained in \cite{luwangsusylif,lupogyraton}.  In this paper,
we included both of the curvature-squared super-invariants, namely the one
based on the square of the Weyl tensor and the one based on the square
of the Ricci scalar.  We found
large classes of domain wall solutions, as well as Lifshitz and
gryating Schr\"odinger vacua.  Amongst these solutions, we found subsets
that were supersymmetric or pseudosupersymmetric.  We also
obtained (pseudo-)supersymmetric solutions that were asymptotic to
the Lifshitz and gyrating Schr\"odinger vacua.  It is worth pointing out
that these supersymmetric solutions depend upon non-trivial contributions
from the auxiliary fields.  Thus the mechanism for supersymmetry in our
solutions is rather different from that in an on-shell theory,
where typically supersymmetry is associated with a balance between mass and
conserved charges carried by form-fields.  In fact, the massive vector that is essential for the supersymmetry has no conserved charge.

   The wealth of supersymmetric vacua of the AdS, Lifshitz and
gyrating Schr\"odinger types leads to many new avenues for investigation
in off-shell higher-derivative supergravities.
They may provide a rich source of gravity backgrounds for studying the
correspondences of both AdS/CFT and AdS/CMT physics.  In particular, 
the existence of supersymmetric Schr\"odinger and 
gyrating Schr\"odinger vacua provides a supersymmetric framework for
studying non-relativistic field theories.

\section*{Acknowledgements}

We are grateful to Ergin Sezgin for useful conversations, and to the
KITPC, Beijing, for hospitality during the course of this work. 
Y.P. thanks Qing-Guo Huang at the ITP,
Beijing, for hospitality.
The research of H.L. is supported in part by NSFC grants 
11175269 and 11235003. C.N.P. is supported in part by
DOE grant DE-FG03-95ER40917.

\end{document}